\documentclass[twocolumn,prl,showpacs,10pt]{revtex4}
\newcommand{\be}{\begin{equation}}
\newcommand{\ee}{\end{equation}}
\begin{document}
\input epsf
\title{Decaying particles and the \\ reionization history of the Universe}
\author{Elena Pierpaoli}
\email{epierpao@princeton.edu}
\affiliation{ Physics Department and Astronomy  Department,
 Princeton University, Princeton, NJ, 08544~~USA}

\date{November 13, 2003}

\begin{abstract}

We investigate the possibility that the Universe is significantly 
reionized by the decay products of heavy particles. 
The ionization produced by 
decay particles implies a high optical depth even if the 
maximum level of ionization ever produced is low ($10^{-2}$). As a 
consequence, a high ionization fraction ($x \simeq 0.5$) at high redshifts 
($z \simeq 20$) fails to fit the cosmic microwave background (CMB) spectra 
at $l \ge 30$. Recent CMB data  limits the primordial abundance of the 
decaying particles, favoring long decay times. 
Other significant sources of reionization are still
needed at $z \simeq 13$. The decay process heats up  the medium, bringing 
the expected $y$ distortion to unobservable levels.


\end{abstract}


\maketitle

\vskip2pc

\renewcommand{\thefootnote}{\arabic{footnote}}
\setcounter{footnote}{0}

\underline{\em 1)  Introduction}

The recent results of the WMAP satellite \cite{Ben03} have evidenced
that the universe presents a higher optical depth that previously
anticipated from the observation of  absorption lines toward quasars
at redshift $z \simeq 6$. The data suggest that the Universe is 
highly ionized at redshifts as high as $z\simeq20$.
Since this discovery, scientists have investigated different possibilities
for producing this high--redshift reionization. 
In the standard reionization scenario where the first stars are 
fully responsible for reionization, the WMAP result would imply 
great difficulties for a number of models, namely the ones in which
the matter power spectrum at small scale is suppressed.
Among these are, for example,
 the warm dark matter (WDM) model of structure formation
\cite{PBMY98,col96} and the cold dark matter model with a running spectral
index. These models wouldn't provide the sufficient
amount of small scale power to produce the needed amount of bound objects
at the appropriate (high) redshifts \cite{bark01}.
The issue is particularly important because the WMAP team data analysis
seem to favor a running spectral index, when no link between reionization
redshift and value of the running spectral index is assumed.

This difficulty may be overcome if the recombination/reionization 
history is altered by some non--standard process, like black hole 
evaporation or the decay of heavy particles, and theorists
have started to develop specific models to this aim \cite{NN02}.
\citet{BMS03} analyzed the impact of the recombination history 
implied by the \cite{NN02} models. As  for decaying particles,
their analysis is limited to fairly long--lived  specific
Supermassive Dark Matter particles.  
In a recent paper, \citet{HH03} (hereafter HH)
 have proposed a model in which the 
products of heavy sterile neutrino decay
 would significantly reionize the Universe at redshift $z\simeq 20$.
At difference with the previous models, these neutrinos may have a shorter
lifetime and an inferred abundance linked to a different physics.
 
Given the apparent variety of decaying particle models, we take here 
a general approach and ask whether current data already put constraints on 
the   reionization history implied by the decay and, in turn, on the
particle physics model involved.
We revisit the reionization history implied by decaying particles 
and compare the implied CMB and large scale structure spectra with 
current data.
We initially perform our calculations in the HH model,
 which we then generalize to the case of unspecified 
decay time and  particle abundance.
We work in a flat Universe with 
$\Omega_m = 0.3$, $\Omega_\Lambda=0.7$, $\Omega_b=0.05$  and $H_0 =70$ Km/s Mpc$^{-1}$.

\underline{\em 2) The decaying particle model for reionization }

We consider here particles with a decay time $t_{dec}$ of the order of 
 $10^{15}~ {\rm s}$
(and define $\tau_{15} \equiv t_{dec}/10^{15} {\rm s}$). The typical decay
redshift for such particles would be $z_{dec} \le 20$.
For clarity, we refer here to the model presented by HH, in which 
a massive neutrino with mass $m_x$ of a few hundred MeV decays into 
an electron plus a pion. The main mechanism of reionization in this case
is the following:
the relativistic electrons produced by the decay process 
 inverse Compton scatter the cosmic microwave background (CMB)
 photons, which then
reionize the hydrogen atoms.
Each electron would be able to cause the reionization of a number of
 photons, according
to its energy $E_e =(M-1) m_\pi/2$ where $M\equiv m_x/m_\pi >1 $.
Typical electron energies are of the order of 1--100 MeV;
each electron is therefore expected to result in
about $10^6$ ionizations,
with an efficiency of about 1/3.
The sterile neutrino comoving  number density 
is (HH): 
$
 n_x \simeq 5.5 \times 10^{-7} {1 \over {\tau_{15}(M^2-1)}}~ m^{-3}, 
$
which would imply an abundance at the present time of:
$
 \Omega_x \simeq 7 \times 10^{-9} {M \over {\tau_{15}(M^2-1)}},
$
 irrelevant for  determining the expansion  at any time \footnote{The sterile
neutrinos are produced in the early Universe when $T \simeq 7 $~GeV
 through neutrino oscillations.
Their number density depends upon their mass and mixing angle. 
Because their decay time also depends on mass and mixing angle, it is possible
to express  $\Omega_x$ as a function of $M$ and $\tau_{15}$.}.

The decay process 
produce an additional ionization source which should be added to the standard
rate in the ionization  equation:
\begin{eqnarray}
{\left({dn \over dt} \right)}_{dec}= \Gamma_x (E_e/I_H) \epsilon = \Gamma_x {E_e \over \Delta E}
\label{eq:io} 
\end{eqnarray}
where $\Gamma_x = dn_x/dt$ in comoving coordinates, $I_H = 13.6~{\rm eV}$
 and the efficiency  $\epsilon$ is approximately 1/3, leading to an energy
 per reionization $\Delta E \simeq 50 ~{\rm eV}$. 

Because only a fraction of the electron energy is used to produce ionization,
the remaining part is assumed to heat the intergalactic medium.
The additional term in the  equation 
for the matter temperature $T_m$ therefore reads:
\be
{\left(d T_m/dt \right)}_{dec}
 = {{2 \Gamma_x E_e} \over {3 n k_B }  } \left(1 - {{I_H + 3/2 k_B T_m} \over 
{(I_H/\epsilon)}} \right)
\label{eq:temp} 
\ee

\begin{figure}[t]
\centering
\hspace*{-4.5mm}
\leavevmode \centerline{\epsfysize=8.0cm \epsfbox{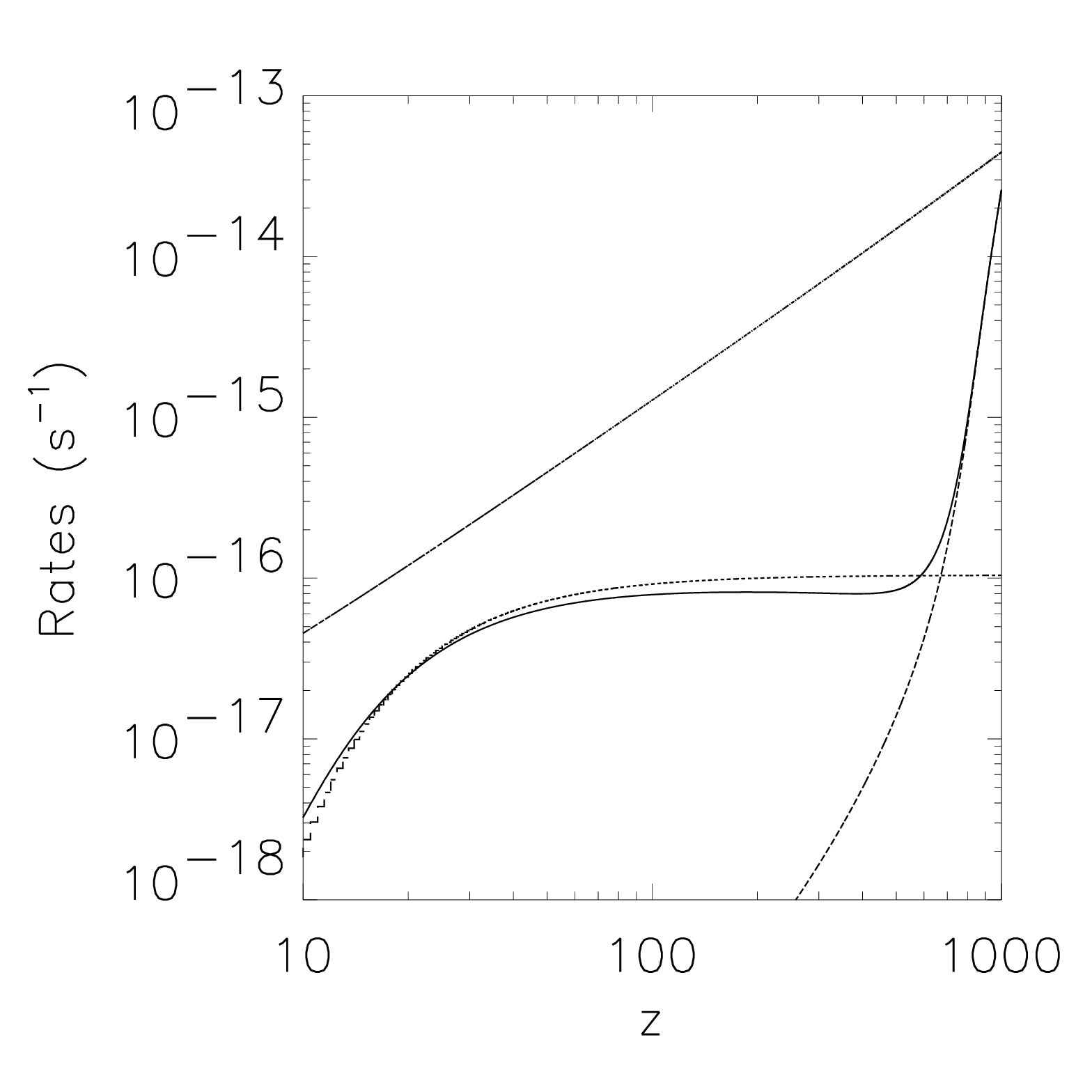} }\\[3mm]
\caption[fig1]{\label{fig1} 
Different rates are plotted as a function of redshift.
Within the  ``standard'' decaying particle model ($m_x=200$ MeV,
$\tau_{15}=4$)),
the solid line is the recombination rate in the decaying particle model,
 the dotted is the ionization
rate due to the decaying particles  and the straight line is the expansion rate
$H(z)$.
The higher ionization produced by the decay particle implies a higher
recombination rate at redshift $z \simeq 10-600$ with respect to 
the standard reionization one (dashed line).
  }
\end{figure}

\begin{figure}[t]
\centering
\hspace*{-4.5mm}
\leavevmode\epsfysize=8.0cm \epsfbox{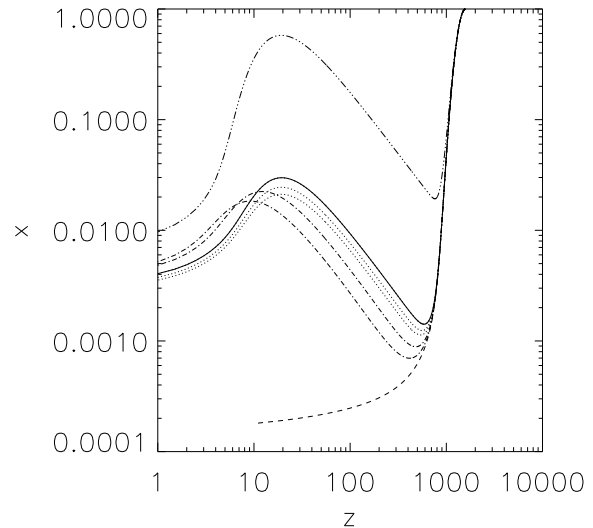}\\[3mm]
\caption[fig2]{\label{fig2} 
Ionization histories  with and without a decay model.
The dashed line correspond to  standard recombination with no reionization.
The solid line refers to the standard decay model adopted here ($m_x=200$ MeV,
$\tau_{15}=4$); note that the maximum reionization implied is $x \simeq 0.03$.
The dot--dashed lines correspond to a decay model with the same particle mass 
but $\tau_{15}=8-12$, while the dotted lines correspond to the same  
$\tau_{15}=4$ and $m_x=350 - 500$ MeV. 
The three dot-dashed line is obtained by the reference model by increasing
the number density of the decaying particle by a factor 300.}
\end{figure}

We modified the RECFAST code \cite{SSS99} by adding the terms in 
eq.~(\ref{eq:io})
and eq.~(\ref{eq:temp}).
Fig.~\ref{fig1} shows  the reionization/recombination rates
for this ionization prescription.
We have chosen as reference parameters $m_x=200$ MeV and $\tau_{15}=4$,
which would imply that most particles have decayed by $z \simeq 20$. 
The presence of the extra-ionization source  boosts the recombination
rate at redshift $z \le 800$ (fig.~\ref{fig1}).
This means that most of the atoms reionized by the decay process
would quickly recombine.
As a result, the maximum ionization fraction produced with this reionization
prescription never exceeds 0.03 (fig.~\ref{fig2}), 
in contrast with what claimed in HH who predict a  50\% ionization 
fraction  $x$ at $z=20$.
Within the sterile neutrino model, it is impossible to reach such level
of reionization just changing the parameters.
A $\tau_{15}$ increase only shifts the maximum  value of $x$ toward lower
redshifts, while altering the mass of the particle cause the neutrino 
number density and electron energy
to have  compensating effects on the ionization fraction.

\begin{figure}[t]
\centering
\hspace*{-4.5mm}
\leavevmode\centerline{
\epsfysize=8.0cm \epsfbox{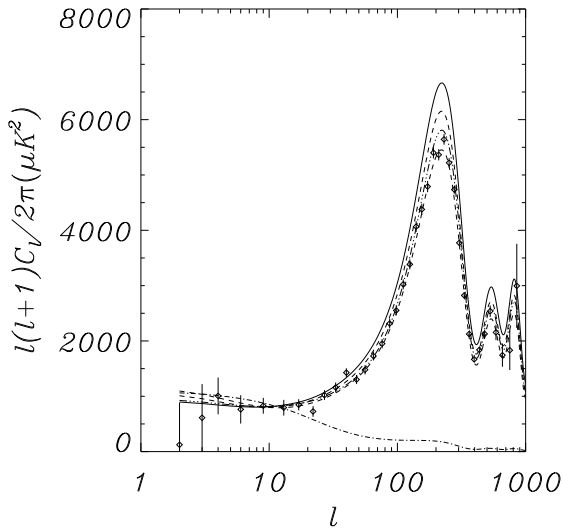}}\\[3mm]
\caption[fig3]{\label{fig3} 
Temperature power spectra of the CMB.
The solid line is the model with no reionization.
The  dashed is a model with sudden reionization at $z\simeq 17$ (implying
the WMAP best fit value $\tau=0.17$).
The three dot--dashed is the standard neutrino decaying model from HH, and 
the dot--dashed line is the decaying model where the abundance of the 
particles have been increased by a factor 300 in order to have $x \simeq 0.5$
at $z=20$.
\label{fig:fig3}

}
\end{figure}

\begin{figure}[t]
\centering
\hspace*{-4.5mm}
\leavevmode\centerline{
\epsfysize=8.0cm \epsfbox{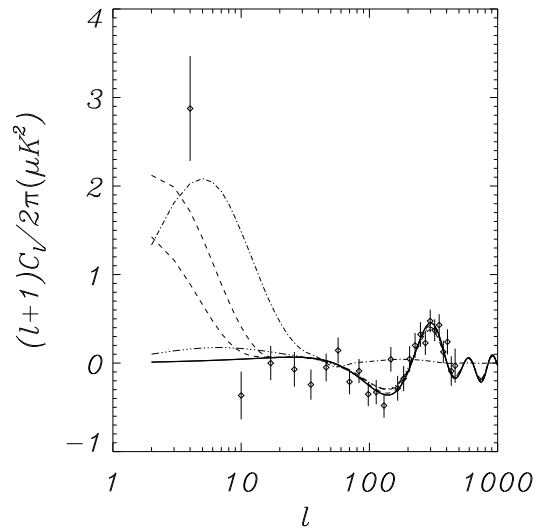}}\\[3mm]
\caption[fig3]{\label{fig4} 
Temperature-Polarization correlation for the models of fig.~\ref{fig:fig3}.
 The lower dashed curve is
a model with sudden reionization at $z=6$, which  falls short in fitting
the low $l$'s power spectrum.

}
\end{figure}

In order to obtain 50\% of reionization at $z=20$ as a
consequence of  the decay process, the number density of the neutrinos
must be artificially increased by a factor  of order 300 (see fig.~\ref{fig2}).
This is not feasible in the sterile neutrino model, but 
it may be conceivable that other particle physics candidates
would produce such abundance.
It is therefore sensible to ask which is the maximum level of reionization
that can be produced by decaying particles without violating 
the actual CMB constraints.
If these particles produce yields which then cause the Universe to reionize,
the reionization process is likely to be similar to the one described
above.
In order to investigate to what extent such reionization process is allowed,
in the following we take  as a reference model  the $m_x = 200~ {\rm MeV}$ 
and $\tau_{15}=4$ sterile  neutrino case to compute a reference abundance
and decay yield energy. 
We then analyze the consequences of an hypothetical particle 
with abundance $f_x$ times  the one of the reference case
 ($\Omega_x = 2.34 \times 10^{-9}$). We consider 
decay times in the range  $1\le \tau_{15} \le 20$.

\underline{\em 3) Implications on the CMB power spectrum  } 
In this section we examine the impact of the modified reionization history
on the CMB power spectrum.
It is well known that reionization has different effects on the 
temperature (TT) power spectrum and on the temperature polarization 
cross--correlation (TE).
In both cases, an increased optical depth $\tau$ causes a damping 
of the spectra that is progressively more pronounced for higher $l$'s.
The TE spectrum, however, is boosted at low $l$'s if  significant reionization 
occurs at low redshift. The earlier reionization occurs, the greater
the $l'$s where the effect appears in the TE spectrum.
While the overall optical depth is to a large extent
 degenerate with other parameters, 
the low--$l$'s bump in the spectra
 is a quite unique signature of reionization, and is precisely
what has been observed by WMAP.
Assuming instantaneous and complete reionization occurring at 
some low redshift $z_{re}$, the WMAP team has 
found $\tau \simeq 0.17$, corresponding to $z_{re} \simeq 17$ \cite{Sper03}.

In the neutrino decay reference model,
the decay process implies an alteration of the reionization history 
already at high redshift ($z \le 800$). 
Despite the fact that the 
 ionization fraction $x$ never reaches 0.1, it remains
 significantly higher than in the standard recombination case for an extended
period of time, during which the number density of baryons   is also high.
This  causes the optical depth to $z\simeq 1000$ to be high
even if the ionization fraction never reaches high values.
For the neutrino decay model with $m_x=200$ MeV and $\tau_{15}=4$, we find
that the optical depth in the range $20 \le z \le 800$ is $\tau=0.11$.
These models therefore provide a way of fitting the TT power spectrum at high 
$l$'s without significantly altering the current accepted ranges
for other parameters (see fig.~\ref{fig3}). 
Note, however, that a significant level of reionization at $z \simeq 10-15$ 
 is needed in order to match the WMAP low--$l$ result of the TE power spectrum,
and sudden reionization at $z=6$ would fall too short in fitting
the low--$l$'s TE spectrum even if the adequate total optical depth was
mainly provided by the decaying particles.
(fig.~\ref{fig4}).
For this reasons, we still have to invoke an early star reionization.

Can decaying particles produce a reionization fraction of $0.5$ at 
$z\simeq 20$ ?
As shown in fig.~\ref{fig2}, in such a reionization scenario the 
ionization fraction would always be $x \ge 0.02$.
The total optical depth would then be extremely high, causing an excessive
damping of the TT and TE power spectra at high $l$'s.
Moreover, the reionization signature in the TE power spectrum would appear
 at too high $l$ values ($l \simeq 30$) contradicting the data.
We conclude that it is not possible to reionize the Universe to high 
levels at $z \simeq 15$ with a  process that is powered by 
dark matter particle decay.

\begin{figure}[t]
\centering
\hspace*{-4.5mm}
\leavevmode\epsfysize=8.5cm \epsfbox{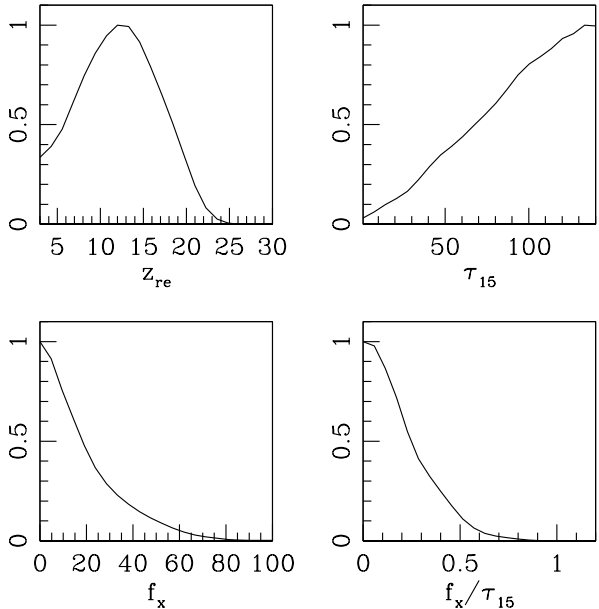}\\[3mm]
\caption[fig5]{\label{fig5} 
Marginalized likelihoods for  the decay model of reionization parameters.}
\end{figure}


\underline{\em 4) Cosmological and astrophysical constraints }
We want to determine to what extent  current cosmological data
allow a particle decay process with 
a decay time shorter than or equal to the age of the Universe to trigger 
reionization.
We performed a multi--parameter fit to the most recent
radiation and matter power spectra data.
We allowed for particle decay reionization with timescales 
$1 \le \tau_{15} \le 140$ 
and arbitrary  abundance (parametrized by $f_x$).
In addition, we allowed for instantaneous reionization at a redshift $z_{re}$.
We kept the spectral index $n_s$ and the amplitude of the matter power
 spectrum $A_s$ as free parameters.  
The results were obtained with a modified version of the CosmoMC package 
\cite{LB02} which uses the WMAP likelihood \cite{V03,H03}.

The results are shown in fig.~\ref{fig5}.
The abundance of decaying particles is limited by $f_x \le 51$
 at 95 \% C.L., 
and there is no evidence in the data that such decay--induced 
reionization models fit the spectra
better than the standard reionization scenario.
Longer lived particles are preferred to short lived ones.
The data naturally constrain the reionization rate in the early 
Universe, which  is proportional to $f_x/\tau_{15}$. Its  95 \% C.L. upper 
limit is  $f_x/\tau_{15} \le 0.48$.
The reason why long--lived particles are favored
resides precisely in the fact that there is a broad range of 
possible abundances for which the reionization history is little affected.
On the contrary, short--lived particles imply a higher reioniation rate,
therefore constraining $f_x$ more.

The marginalized likelihoods show a peak at a  slightly lower reionization 
redshift  ($z_{re}=13$) and a smaller 
optical depth 
with respect to the standard reionization case investigated by the WMAP team, 
but the likelihoods are still wide.
The spectral index is not significantly affected.

As for other astrophysical constraints, 
it has been argued (HH) that decay--induced reionization may produce 
a specific distortion in the cosmic microwave background.
We computed the evolution of matter temperature with the  extra--term
 in eq.~\ref{eq:temp}.
In the decay particle scenario the additional ionization causes 
the matter temperature  to follow the radiation one
for a longer time.
The derived $y$ parameter, which depends on the difference between matter
and radiation temperatures in the past, is therefore very small.
For the reference neutrino model, it is $|y| \le 3 \times 10^{-10}$,  
and increasing the decay particle abundance by a factor 300 
we have $|y| \le 4 \times 10^{-10}$; still too small to be detected.


\underline{\em 5)  Conclusions}
In this paper we analyzed to which extent decaying particles may contribute
to produce the high ionization rate observed at high redshift by the WMAP
experiment.

We showed that if  decay yields had to produce a significant reionization
at redshifts $z \simeq 20$, they
  would alter the reionization history
already at redshift $z \simeq 800$, significantly modifying the 
reionization/recombination rate and the derived ionization fraction at 
all redshifts.
The implied  optical depth would be too high to match the one derived 
from  recent CMB observations.
Therefore the decay mechanism cannot be invoked  to 
produce a significant ionization fraction at high redshift.

As a consequence, a standard reionization mechanism (e.g. starlight) acting at 
fairly high redshifts ($z_{re} \simeq 13$) must still 
be invoked in order to fit the polarization data. 
These findings suggest that having some reionization produced by 
decaying particles wouldn't ease the constraints put on small scale 
matter power spectrum by the requirement of having enough collapsed structure
at high redshifts to produce small scale objects.
Decaying particles do no alleviate the challenge that current data
impose to 
models of structure formation like WDM or CDM with a running spectral index.

We showed that the current CMB and large scale structure data 
constrain the abundance of the decaying 
 dark matter particles to 
 $\Omega_x \le 1.3  \times 10^{-7} (1+z)^3$ at 95 \% C.L., 
 when the decay time is  in the range $1-140 \times 10^{15}$~s.
Within this range, the data tend to prefer long--lived particles.
This result translate in a limit on the fraction of electrons reionized 
per unit time in the early Universe: $2 \times 10^{-16}~{\rm s}^{-1}$. 
There is no evidence that this altered  reionization prescription 
produce a better fit to the data  than the instantaneous reionization one.
We evaluated the $|y|$ distortion that would be expected from this 
reionization mechanism, 
and found it to be below $10^{-9}$, therefore too small to be observed.

\section*{Acknowledgments}
E.P. is supported by NASA grant NAG5-11489. She wishes to thank Bruce Draine
for several helpful discussions, the Aspen
Center for Physics  and the University of Cagliari (Italy) for hospitality 
during the preparation of this work.

\bibliography{reio}

\end{document}